\documentclass[aps,pre,twocolumn,groupedaddress]{revtex4-1}
\usepackage{graphicx}

\begin{document}

\title{Impact of intrinsic biophysical diversity on the activity of spiking neurons}

\author{Man Yi Yim, Ad Aertsen, Stefan Rotter}
\email[]{stefan.rotter@bcf.uni-freiburg.de}
\affiliation{Bernstein Center Freiburg \& Faculty of Biology, University of Freiburg, Freiburg, Germany}

\date{\today}

\begin{abstract}
We study the effect of intrinsic heterogeneity on the activity of a population of leaky integrate-and-fire
neurons. By rescaling the dynamical equation, we derive mathematical relations between multiple neuronal parameters and a fluctuating input noise. To this end, common input to heterogeneous neurons is conceived as an identical noise with neuron-specific mean and variance. As a consequence, the neuronal output rates can differ considerably, and their relative spike timing becomes desynchronized. This theory can quantitatively explain some recent experimental findings.
\end{abstract}


\maketitle

\section{Introduction}
\begin{figure}
\includegraphics[scale=0.22]{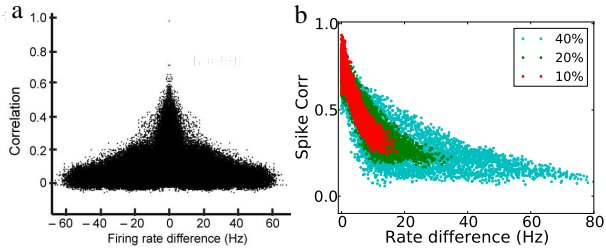}
\caption{(Color online). (a) Correlation coefficient ($1\,\mathrm{ms}$ window) of $589$ spike trains from mitral cells \textit{in vitro} receiving an identical input as a function of rate difference, adapted with permission from \cite{Padmanabhan2010}.
(b) The same measure for $100$ simulated heterogeneous LIF neurons using uniformly distributed $\tau_{m}$, $\tau_{r}$, $C$ and $\theta$ with different distribution widths.
\label{f1}}
\end{figure}

In statistical physics, it is often assumed that individuals are intrinsically identical. In neuroscience also, identical parameters are typically assumed for all neurons in the study of neuronal population activity and correlation transmission. Real neurons, though, even if they are of the same type and located in the same brain area, exhibit intrinsic differences. Their morphologies and the intracellular concentrations of ions, to name just two examples, can differ widely, although in principle they have been generated by the same mechanisms \cite{Marder2006}. As a consequence, neuronal spike patterns can differ although neurons receive identical inputs \cite{Mainen1995,Padmanabhan2010}. Recently, \textit{in vitro} intracellular recordings of isolated mitral cells in the mouse olfactory bulb were conducted while they responded to identical input \cite{Padmanabhan2010} (Fig.\,\ref{f1}(a)). The neurons displayed diverse output firing rates and pairwise correlations. Specifically, the spike correlation coefficient obtained with a $1\,\mathrm{ms}$ observation window co-varied with the rate difference of the neuron pairs: small differences resulted in a wide range of different spike correlations, but large differences led always to small spike correlation.

In homogeneous network models, additional independent Gaussian white noises or independent Poisson spikes are very often added to every constituent identical neuron to account for their diverse spike timing. In real brain networks, not only the spike timing but also the spiking rate of neurons differ due to their intrinsic biophysical diversity. Therefore, it is of great interest to understand how the biophysical heterogeneity of a neuronal population contributes to neural coding and information processing in neuronal networks. Research work has been conducted on the coding properties \cite{Chelaru2008,Mejias2012} and synchronous responses \cite{Tsodyks1993,Wang1996,Brette2012} in a network of heterogeneous neurons. In many cases, neuronal heterogeneity was implemented simply by replacing one or more fixed neuronal parameters, such as the offset current \cite{Tsodyks1993,Wang1996}, the spiking threshold \cite{Mejias2012}, or the synaptic conductance \cite{Chelaru2008}, by a Gaussian- or uniformly-distributed random variable.

Here we investigated more fundamental questions, using both theoretical analysis and simulations: how neuronal heterogeneity can be represented appropriately in theory and how it can affect the neuronal dynamics and the spiking statistics in a population of simple leaky integrate-and-fire (LIF) neurons. The limitations of the existing approaches are addressed first. Then we suggest a more general scheme to implement biophysical diversity when either rate or correlation is of interest. By rescaling the dynamical equation, we derive mathematical relations between multiple neuronal parameters and the input noise. The main impact of common input to heterogeneous neurons on rate and correlation can be realized by an identical (frozen) noise current injection with different values of mean and variance, whereas the complete effect is captured by additionally drawing distributed values of the membrane time constant and the refractory period. In this scheme, the rate difference of heterogeneous LIF neurons can be treated analytically. As for correlation, we utilize alternative correlation measures to illustrate that spikes from heterogeneous neurons may be desynchronized by several milliseconds, thus escaping detection by a $1\,\mathrm{ms}$ observation window.

\section{Model}
We consider a population of isolated leaky integrate-and-fire (LIF) neurons, each of which has its membrane potential $V(t)$ governed by
\begin{eqnarray}
 \tau_{m}\dot{V}(t) &=& -V(t) + RI(t), \label{lif}
\end{eqnarray}
where the input synaptic current
\begin{eqnarray}
RI(t) &=& \tau_{m}J_{E}\sum_{j}\delta(t-t_{j}) - \tau_{m}J_{I}\sum_{k}\delta(t-t_{k}).
\end{eqnarray}
$\tau_{m} = RC$ is the membrane time constant. $R$ and $C$ are the membrane resistance and capacitance, respectively. $J_{E}$ ($J_{I}$) is the amplitude of an excitatory (inhibitory) post-synaptic potential, whereas $t_{j}$ ($t_{k}$) represents the time of the $j$th ($k$th) excitatory (inhibitory) input spike.
When $V(t) = \theta$, $V(t)$ is reset to $V_{r}$ and a pause for synaptic integration $\tau_{r}$ is imposed to mimic the refractory period. In the high-input regime, the sum of synaptic inputs to a neuron can be approximated by a fluctuating input noise \cite{Ricciardi1979,Kuhn2004}
\begin{eqnarray}
 I(t) &\equiv& \tau_{m}[\mu + \sigma\eta(t)], \label{current}
\end{eqnarray}
where
\begin{eqnarray}
 \mu &=& J_{E}\nu_{E} - J_{I}\nu_{I}, \label{mu} \\
 \sigma &=& \sqrt{J_{E}^{2}\nu_{E} + J_{I}^{2}\nu_{I}}. \label{sigma}
\end{eqnarray}
$\eta(t)$ is a white noise random process such that $\langle \eta(t)\eta(t')\rangle = \delta(t-t')$. $\nu_{E}$ ($\nu_{I}$) is the firing rate of the excitatory (inhibitory) input.

The numerical integration of Eq.~\ref{lif} in our simulations was performed using the fourth-order Runge-Kutta method with a time step of $0.01\,\mathrm{ms}$.

\section{Heterogeneity}
Intrinsic diversity of a population of neurons can be directly imposed by drawing neuronal parameters from a distribution. Here, we tested the response of $100$ isolated heterogeneous LIF neurons to an identical fluctuating input current in the form of Eq.\,(\ref{current}). Neuronal heterogeneity is implemented by drawing four uniformly distributed parameters: $\tau_{m}$, $\tau_{r}$, $C$ (or $R$) and $\theta$, which, together with $V_{r}$, represent all the independent parameters of a LIF neuron in response to a current input. The mean values of the uniform distribution are $20\,\mathrm{ms}$, $2\,\mathrm{ms}$, $1$ and $1$ (in arbitrary units) respectively, and $V_{r}$ is fixed to $0$. These values are used throughout this work. We maintain the temporal scale of the dynamics of a typical neuron and rescale the potential by setting the mean reset to zero and the mean threshold to $1$. The correlation coefficient as a function of the output firing rate difference of all possible pairs with different distribution widths (percentage with respect to the mean) from $100\,\mathrm{s}$ of simulations (an example with $\mu = 0.03$ and $\sigma = 0.3$ shown in Fig.\,\ref{f1}(b)) highly resemble the experimental findings in \cite{Padmanabhan2010} (Fig.\,\ref{f1}(a)).

\begin{figure}
\includegraphics[scale=0.15]{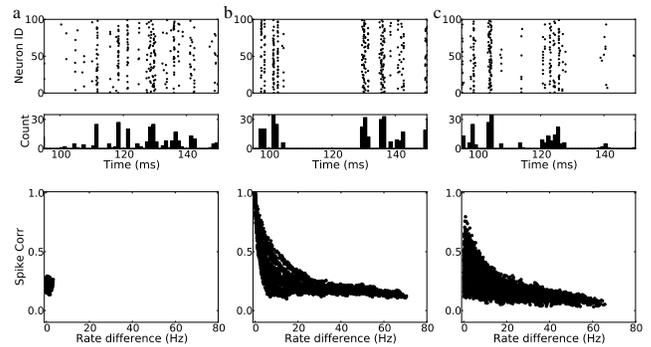}
\caption{Rasterplot, peri-stimulus time histogram (PSTH) and correlation coefficient as a function of rate difference for the following input into $100$ LIF neurons:
(a) $\mu + \sigma\big[\sqrt{c}\eta(t) + \sqrt{1-c}\xi_{i}(t)\big]$ ($c = 0.9$); 
(b) identical input $\mu + \sigma\eta(t)$ but distributed $\theta_{i}$;
(c) our scheme using $\mu_{i} + \sigma_{i}\eta(t)$ together with distributed $\tau_{m}$ and $\tau_{r}$, consistent with both the experimental findings \cite{Padmanabhan2010} and the mathematical analysis. The subscript $i$ denotes ``independent''.
\label{f2}}
\end{figure}

Diverse neuronal spike timing in a network has very often been achieved by adding independent random inputs to individual neurons. We provide every identical neuron with a common input as the input signal plus an independent input with the same statistics among neurons , in the form of
$\mu + \sigma\big[\sqrt{c}\eta(t) + \sqrt{1-c}\xi_{i}(t)\big]$ where $\eta(t)$ and $\xi_{i}(t)$ are independent Gaussian white noises. Fig.\,\ref{f2}(a) displays the raster and the correlation coefficient as functions of the rate difference when $\mu = 0.06$, $\sigma = 0.2$ and $c = 0.9$. The rate difference is close to zero and the correlation coefficient between any pair is nearly the same \cite{delaRocha2007}. Decreasing $c$ leads to a drop in spike correlation but has no effect on the rate difference. These observations are very distinct from both the experimental (Fig.\,\ref{f1}(a)) and simulation (Fig.\,\ref{f1}(b)) results. In view of some previous work on the reliability of single neurons in response to a repeated input \cite{Mainen1995,Teramae2008,Padmanabhan2010}, this implementation may be adopted to account for trial-to-trial variability of the same neuron.

It is common practice to implement heterogeneity of neurons by drawing random variables for a single neuronal parameter. To test its validity, we provide every identical neuron with a common input plus a random value of the spiking threshold drawn from a uniform distribution $\theta \in [0.5,1.5]$. Fig.\,\ref{f2}(b) displays the raster and the correlation coefficient as functions of the rate difference. Another example with distributed values of the input offset current $\mu$ instead of  $\theta$ is shown in Fig.\,\ref{f3}(a). In either case, neurons exhibit dispersive firing rates. However, the spike correlation distribution at different values of rate difference is too narrow compared with Fig.\,\ref{f1}(a) and (b), and the region of small rate difference and small spike correlation cannot be reached. Thus, the impact of neuronal heterogeneity is only partially accounted for. We describe our alternative approach in the next section.

\section{Mathematical relations}
In addition to the five parameters mentioned above, two additional parameters correspond to synaptic inputs: $J_{E}$ and $J_{I}$. We analyze the contribution of the heterogeneity in the seven independent neuronal parameters in a population of LIF neurons in the high-input regime.

Regarding synaptic inputs, heterogeneity in the parameters $J_{E}$ and $J_{I}$ can be captured by heterogeneity in $\mu$ and $\sigma$, according to Eqs. (\ref{mu}) and (\ref{sigma}). For example, if $J_{E}$ is uniformly distributed, $\mu$ is also uniformly distributed, whereas $\sigma^{2}$ is distributed as the square of a uniformly distributed variable.

The other five parameters are present in the neuronal dynamics irrespective of the type of inputs (current or spikes). First, $R$ (or $C$, depending on the form of writing) can be absorbed into $I(t)$ as shown in Eq.\,(\ref{lif}) so any distribution of $R$ can be accounted for by a corresponding distribution of $\mu$ and $\sigma$.

The difference between $\theta$ and $V_{r}$, which is the potential difference a neuron has to traverse, is a quantity relative to the synaptic strengths $J_{E}$ and $J_{I}$. For instance, lifting $\theta$, or lowering $V_{r}$, is equivalent to reducing $J_{E}$ and $J_{I}$ together by the same ratio. Thus, heterogeneity in $\theta$ and $V_{r}$ can be included in $\mu$ and $\sigma$ by means of rescaling.

Unlike the above five parameters related to the potential, the remaining two shaping the neuronal response in the temporal scale, $\tau_{m}$ and $\tau_{r}$, cannot be rescaled or captured by $\mu$ and $\sigma$. Their distributions among neurons have to be accounted for separately. Therefore, in the high input regime when the approximation of a fluctuating input noise is valid, the seven independent neuronal parameters (and their distributions) can be reduced to four: $\mu$, $\sigma$, $\tau_{m}$ and $\tau_{r}$. Based on this analysis, we suggest using distributed values of these four parameters together with an identical noise $\eta(t)$ to account for the effects of all the parameters in a population of LIF neurons receiving identical inputs. This is in contrast to the common practice of using independent noises as shown in Fig.\,\ref{f2}(a). We draw the parameters from uniform distributions 
$\mu \in [0.015,0.105]$, $\sigma \in [0.1,0.3]$, $\tau_{m} \in [16,24]\,\mathrm{ms}$ and $\tau_{r} \in [1.5,2.5]\,\mathrm{ms}$. 
In Fig.\,\ref{f2}(c), both the rate difference and the correlation coefficient, as well as their relation, are consistent with the experimental and our simulation results.

\begin{figure}
\includegraphics[scale=0.2]{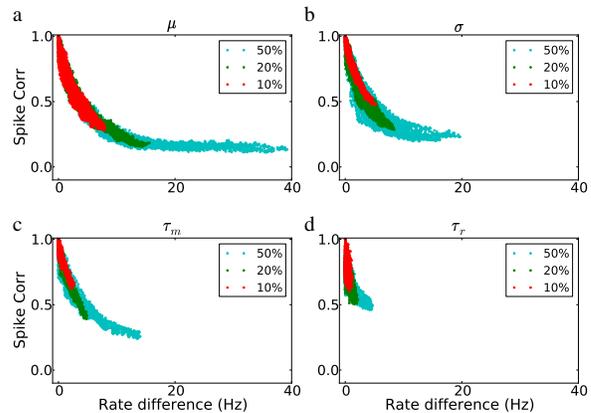}%
\caption{(Color online). Correlation coefficient as a function of rate difference when neurons receive identical noise with uniformly distributed (a) $\mu$, (b) $\sigma$, (c) $\tau_{m}$ and (d) $\tau_{r}$ separately. \label{f3}}
\end{figure}

The respective contributions of the four parameters are investigated. Fig.\,\ref{f3} shows the correlation as functions of the rate difference for $\mu$, $\sigma$, $\tau_{m}$ and $\tau_{r}$ separately. Each realization is drawn from a uniform distribution of  $10\,\%$, $20\,\%$ and $50\,\%$ around their mean values, which are $0.06$, $0.2$, $20\,\mathrm{ms}$ and $2\,\mathrm{ms}$, respectively. The firing rate of a neuron is largely shaped by $\mu$, whereas the distributed values of the variance give rise to different degrees of imprecise spiking. The wider the distribution of $\tau_{m}$ and $\tau_{r}$, the larger is the rate difference and the lower the correlation. When $\tau_{r}\,\ll\,1/\nu$, the effect of $\tau_{r}$ is small.

\section{Rate difference}
\begin{figure}
\includegraphics[scale=0.15]{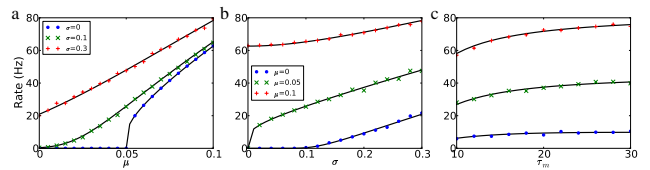}%
\caption{(Color online). Output rate as a function of (a) $\mu$, (b) $\sigma$ and (c) $\tau_{m}$ from theory (black) and simulation (colored).
\label{f4}}
\end{figure}

The firing rate of a LIF receiving a Gaussian distributed noise is known analytically \cite{Brunel1999,Siegert1951}:
\begin{eqnarray}
 \frac{1}{\nu} &=& \tau_{r} + \tau_{m}\sqrt{\pi}\int_{\frac{V_{r}-\mu\tau_{m}}{\sigma\sqrt{\tau_{m}}}}^{\frac{\theta-\mu\tau_{m}}{\sigma\sqrt{\tau_{m}}}}due^{u^{2}}\big(1+\text{erf}(u)\big), \text{when } \sigma > 0; \nonumber\\
 \frac{1}{\nu} &=& \tau_{r} - \tau_{m}\text{ln}(1-\frac{\theta}{\mu\tau_{m}}), \text{when } \sigma = 0.
\end{eqnarray}
Only six parameters $\mu$, $\sigma$, $\tau_m$, $\tau_r$, $\theta$ and $V_{r}$ influence the firing rates, of which only the first four are independent. Changing any of them can result in a rate difference as shown in Fig.\,\ref{f4}, and this explains the wide distribution of firing rates in a heterogeneous neuronal population.

\section{Imprecise spiking}
\begin{figure}
\includegraphics[scale=0.15]{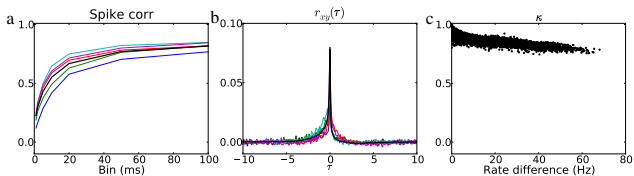}%
\caption{(Color online). 
(a) Correlation coefficient as a function of window size (colored lines indicate five examples; the black thick line indicates the average over all pairs).
(b) Cross-correlation. Positive correlation at negative $\tau$ indicates that the higher-rate neuron is leading.
(c) $\kappa$, which takes jittered spikes into account, as a function of rate difference.
\label{f5}}
\end{figure}

The raster plot in Fig.\,\ref{f2}(c) shows population synchrony with spike-time jitter. Low average correlation coefficients in a population of neurons do not necessarily imply an asynchronous state of the population. Removing spikes from one of the two identical spike trains can also reduce the correlation coefficient significantly. When we compute the correlation coefficient in our data with a larger bin size, larger values for the the correlation coefficient are obtained, as shown in Fig.\,\ref{f5}(a). This is because some spikes become ``coincident'' only for larger bin sizes. The output spike trains behave like jittered spikes as discussed in earlier theoretical studies \cite{Tetzlaff2008}. Whether the decorrelation due to neuronal heterogeneity is significant depends critically on the bin size, or the integration window of the neurons receiving such inputs.

In view of a significant number of spikes jittering outside a $1\,\mathrm{ms}$ bin, we look into the cross-correlation function
\begin{eqnarray}
 r_{xy}(\tau) &=& \frac{c_{xy}(\tau)}{\sigma_{x}\sigma_{y}} = \frac{\langle x(t)y(t+\tau)\rangle - \langle x(t)\rangle\langle y(t)\rangle}{\sigma_{x}\sigma_{y}},
\end{eqnarray}
where $x(t)$ and $y(t)$ denote two output spike trains in discrete time, consisting of 0 and 1 with bin size of $0.1\,\mathrm{ms}$. $x(t)$ is assigned to be the spike train with lower spike count. $c_{xy}(\tau)$ is the covariance function and $\sigma_{x}$ and $\sigma_{y}$ denote the standard deviation of the two spike trains, considered as discrete signals. Fig.\,\ref{f5}(b) shows that the mean $r_{xy}(\tau)$ over all pairs is positive in a small neighborhood of $\tau$, indicating a higher than chance level to observe spikes. Spikes are jittered, instead of asynchronous. In addition, $r_{xy}(\tau)$ is asymmetric, and its area is skewed towards negative $\tau$, indicating that the higher-firing-rate neuron is more likely to lead in terms of spiking \cite{Ostojic2009,Tchumatchenko2010}. 

We further look at the normalized total cross-covariance $\kappa$ \cite{Bair2001,Tetzlaff2008}
\begin{eqnarray} 
\kappa &=& \frac{\int_{-\infty}^{\infty}d\tau c_{xy}(\tau)}{\sqrt{\int_{-\infty}^{\infty}d\tau c_{xx}(\tau)\int_{-\infty}^{\infty}d\tau' c_{yy}(\tau')}},
\end{eqnarray}
which is an overall measure for the fraction of the spikes that are correlated above chance level. Fig.\,\ref{f5}(c) shows that $\kappa$ is quite close to unity, and has a weak dependence on the rate difference. This indicates that spikes would not be decorrelated when a larger time window is considered.

\section{Discussion}
We remark that the plots of the correlation coefficient as a function of rate difference from our simulations (such as Figs. \ref{f1}(b) and \ref{f2}(c) can fit the experimental results in \cite{Padmanabhan2010} more satisfactorily by introducing a random delay of up to $1\,\mathrm{ms}$ to every incoming input spike. The spike correlation distribution then becomes broader, and the regime of low correlated output at small rate differences can also be reached (data not shown). This effect could be due to some (unknown) biological mechanism not captured in a simple LIF neuron model.

We emphasize that common input into heterogeneous neurons is better realized by a shared noise with distributed mean and variance and, more completely, with additionally distributed values of the membrane time constant and the refractory period. This insight is based on both the mathematical analysis presented here and the \textit{in vitro} experimental findings \cite{Padmanabhan2010}. As far as firing rates and spike correlations are concerned, the distributions of mean and variance of the input account for most of the experimental observations. Diversity of $\tau_{m}$ and $\tau_{r}$ has a smaller effect on the quantities in question, nevertheless, including them can account for the full degree of heterogeneity.

In the raster plot of a neuronal population with similar rate differences and spike correlations, synchrony is obvious, although spike times are not precise. Spikes, if present, are jittered in the millisecond range, which cannot be captured by the $1\,\mathrm{ms}$ temporal window used for analysis. This is why using larger bins leads to larger values for spike correlation. We emphasize that neuronal heterogeneity alone does impose an appreciable decorrelation effect on the population activity. However, whether decorrelation is functionally significant depends on the readout of the downstream neurons. On top of that, a network of heterogeneous neurons may give rise to richer network dynamics. It remains to be explored whether such intrinsic heterogeneity can facilitate other decorrelation mechanisms to increase the amount of information flow \cite{Renart2010,Wiechert2010,Padmanabhan2010}. Given the significant reduction in spike correlation among heterogeneous neurons, research concerning correlation transmission must take neuronal heterogeneities into consideration.

\begin{acknowledgments}
We thank Volker Pernice, Moritz Deger, and Tom Tetzlaff for discussions. The present work was supported by the German Federal Ministry of Education and Research (BMBF Grant No. 01GQ0420 to ``BCCN Freiburg'' and BMBF Grant No. 01GW0730 ``Impulse Control'') and the EU (INTERREG-V Grant to Neurex: TriNeuron).
\end{acknowledgments}

\bibliography{corr}

\end{document}